# From Trend Analysis to Virtual World System Design Requirement Satisfaction Study

Bingqing Shen,[1] Weiming Tan,[2] Jingzhi Guo,[2] Hongming Cai,[1] Bin Wang,[2] and Shuaihe Zhuo[3]

[1] School of Software, Shanghai Jiao Tong University, China.

[2] Faculty of Science and Technology, University of Macau, Macau, China.

[3] School of Business, Macau University of Science and Technology, Macau, China.

## Abstract

Virtual worlds have become global platforms connecting millions of people and containing various technologies. The development of technology, shift of market value, and change of user preference shape the features of virtual worlds. In this paper, we first study the new features of virtual worlds and emergent requirements of system development through trend analysis. Based on the trend analysis, we construct the new design requirement space. We then discuss the requirement satisfaction of existing virtual world system architectures and highlight their limitations through a literature study. The comparison of existing system architectures sheds some light on future virtual world system development to match the changing trends of the user market. At the end of this study, a new architecture from an ongoing research, called Virtual Net, is briefly discussed on its possibility in requirement satisfaction.

## 1. Introduction

Virtual worlds have been evolved from simple text-based games to sophisticated three-dimensional graphic environments, covering both game worlds [1] and social worlds [2]. In this evolution, they have developed some important features such as immersion and persistency [3]. Supported by these features, the growth of virtual worlds has brought millions of people to participate and innovation in many fields, including education, medicine, tourism, commerce, and entertainment. As can be expected, they will play an important role in the future economy and society.

Virtual worlds are global computing infrastructure. They are synthetic, sustainable, and immersive environments, facilitated by computer networks connecting multiple users in the form of avatars, who interact in real-time [3], [4]. The evolution of virtual worlds is closely related to their underlying technologies. Virtual world related technologies, including hardware and network, graphics, artificial intelligence, peripherals, human interface, etc., have a wide range over the whole information and communications technology (ICT) sector. These technologies are also evolving and upgrading over the years, spirally changing virtual worlds and bringing new features. Based on technology innovation, this paper identifies the impetus which advances virtual world development to quest for architectural reflection and transformation.

This paper first answers the question of what changes virtual world with a trend analysis. The trends of related technologies are thoroughly studied for eliciting new design requirements, which compliments existing requirements of consistency, scalability and security. Following the trends and their implications, new features of virtual worlds are suggested and summarized as emergent design requirements: sufficiency, reliability, persistency, and credibility. Then, the existing virtual world architectures are scrutinized based on the emergent requirements. By classifying and comparing them in requirement satisfaction, we find that none of the existing system architectures can fully satisfy all requirements due to their intrinsic limitations. To light the future, we briefly introduce an ongoing research, a new distributed architecture which leverages the advantages of existing architectures at the end for the purpose to elicit new discussions on virtual world system design and development.

This paper has made two contributions. Firstly, it has conducted a trend analysis on the virtual world related technologies. Based on the implication of trend analysis, it has explicated the emergent design requirements, as well as the commonly known requirements for completeness, which can be treated as a roadmap for system design. Secondly, it has thoroughly surveyed the existing system architectures, discussed their satisfaction and limitations on the design requirements, and compared their competence on requirement satisfaction.

The remainder of the paper is organized as follows. Section 2 analyses the trends of virtual world related technologies and their implications. Section 3 discusses the virtual world design requirement space based on the trend analysis. Section 4 studies and compare the existing virtual world system architectures on requirement satisfaction. Section 5 compares the existing architectures. Finally, Section 6 summarizes the paper and introduces the future challenges of a new architecture design.

## 2. Trends Analysis

The latest trends of virtual world related technologies can provide the evidences for studying virtual world evolution. To obtain some rudimentary knowledge of such trends, we conducted a search on Statista[1] with the keyword "Virtual World", which returned 821 results, including statistics, forecasts, and studies. The search results were then examined by relevance and filtered by year range from 2015 to 2035. This reduced the results to 538. We then tagged them with relevant topics and created the tag cloud, as shown in Figure 1. With the obtained knowledge, we further looked into each scope and investigated the trends of related technologies, including mobile computing, social networks, virtual/augmented reality, virtual world applications, internet of things (IoT), wearable devices, game intelligence, multisensory multimedia, and computer graphics. The trend is analysed along five directions: mobility, diversity, interconnectivity, immersion, and intelligence. The representative market/industrial evidence, together with academic/online reports, are elaborated in the sub-sections below, followed by the implications to future virtual world design.

---

[1] Statista is an online portal providing more than 1 million statistics collected from more than 22,500 trusted partners, such as Experian Simmons and Euromonitor. It maintains sources transparency and adherence to academic standards: https://www.statista.com/sources/1/

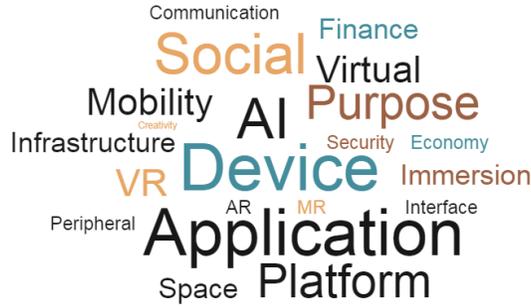

Figure 1 Tag cloud of the most frequently used terms for virtual world.

## 2.1. Mobility

The wide usage of smart devices (e.g., smartphones and tablets) and the pervasion of mobile networks (e.g., the 5th Generation cellular mobile network) promote the concept of playing everywhere and facilitate the development of pervasive gaming [5]. Figure 2 shows the trend of mobile virtual world from 2014. By the end of 2017, the number of mobile gamers has been more than 1.4 times the number of 2014 (Figure 2 (a)). Figure 2 (b) shows that worldwide mobile social gaming revenue has increased by 2.82 billion dollars from 2010 to 2014. Interestingly, the trend of gamer population by device shows that smartphones gain larger traction than tablets. It is likely that, with the increase of smartphone size and capacity, the distinction of smartphone and tablet are diminishing, which reduces the demand on tablet. Moreover, the investigations in year 2017 and 2018 show that smartphone-based applications gains the largest market share both in video game (63%) [6] and virtual reality/augmented reality (77%) [7], compared with PC, laptop, game console, and other standalone game device counterparts. In summary, the trends in mobility show that the user population and consumption on mobile virtual world platforms are growing.

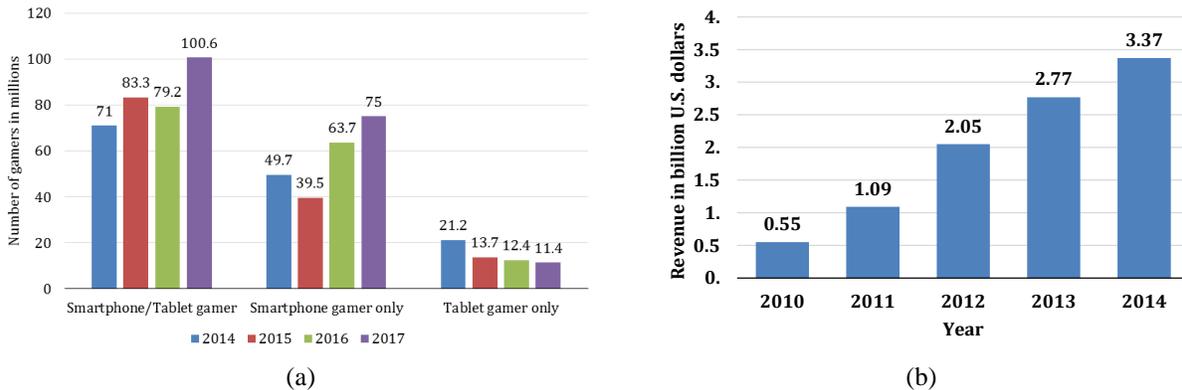

(a) (b)

Figure 2. Trends of mobility: (a) Number of mobile gamers in North America from 2014 to 2017 [8]. (b) Worldwide mobile social gaming revenue from 2010 to 2014 [9].

## 2.2. Diversity

Virtual worlds have been designed to serve different purposes. Figure 3 shows the increasing range of investment from 2016 to 2018. A shift of investment from traditional sectors (gaming, marketing, and military simulation) to new sectors (retail and manufacturing) can be observed. To better satisfy application requirements, various user contents have been generated, such as the virtual objects in Second Life and Minecraft (minecraft.net), to facilitate user's innovation and collaboration [10]. In education, for example, immersive learning environments provide simulated and controlled situations to improve student's performance and motivation in study. Also, dynamic content generation can decrease development effort for teachers to customize learning games [11]. In medicine, the realistic models created in virtual environments can facilitate surgical training [12]. In tourism, virtual world technologies can convert heritage relics to digital assets, which can avoid erosion in long-term preservation [13]. In e-commerce, user involvement in product co-creation [14] can improve product satisfaction.

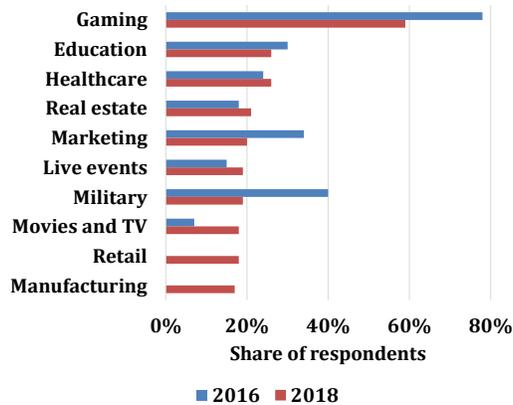

Figure 3. Investment directions of virtual/augmented reality technology worldwide in 2016 and 2018 [15].

## 2.3. Interconnectivity

Virtual worlds can connect different people and devices to support various interaction and communication. Multiple access devices lead to various user data input to virtual worlds. One emergent category is wearable device. Figure 4 (a) shows that user consumption on wearable device is growing in all categories. Wearable devices can collect user data by tracing various user interaction patterns in play. For instance, Cyberith Virtualizer (cyberith.com) can track user posture in walking, running, standing, crouching, and sitting. Data is also collected from IoT devices, creating a mirrored world to telepresent real-world data in a virtual environment. For example, Eye Create Worlds [16] visualizes the data in a virtual world from the sensors placed over a city to monitor and optimize the performance of a rail transportation network. Besides connecting to devices, people in virtual worlds are increasingly inter-connected. Figure 4 (b) shows that the market value of social worlds in 2020 is expected to be twice the value in 2011. Moreover, a game industry survey in 2018 shows 54% gamers players feel video games help them connect with friends and spend time with family [17]. In summary, the above observation shows the interconnectivity trend from both hardware peripherals and social connection, leading to the growth of in-world user data collection.

## 2.4. Immersion

User's experience of immersion is an important trend to study, for it generates a sense of presence [18]. By observing the evolution of the multimedia and computer graphics technology, two trends can be found. First, the market revenue in Figure 5 (a) shows the increasing trend of immersive hardware technology adoption. Second, the recent innovation in multiple sensorial media (mulsemedia) [19] provides users more sensorial experience than visual and audio delight, leading to the deeper experience of immersion. Figure 5 (b) shows the trend of new mulsemedia device release in the haptic rendering, olfactory rending, and gustatory rendering market. The figure for all device categories is growing, especially from 2012. Some VR headsets integrating the multisensory function can provide multiple sensorial experiences (besides audio and video experience). For example, OhRoma (camsoda.com/products/ohroma/) provides different smells during video or audio playing. Feelreal (feelreal.com) can simulate both smell and wind by changing airflow. These trends show the increasing adoption of immersive rendering techniques, enriching user experience.

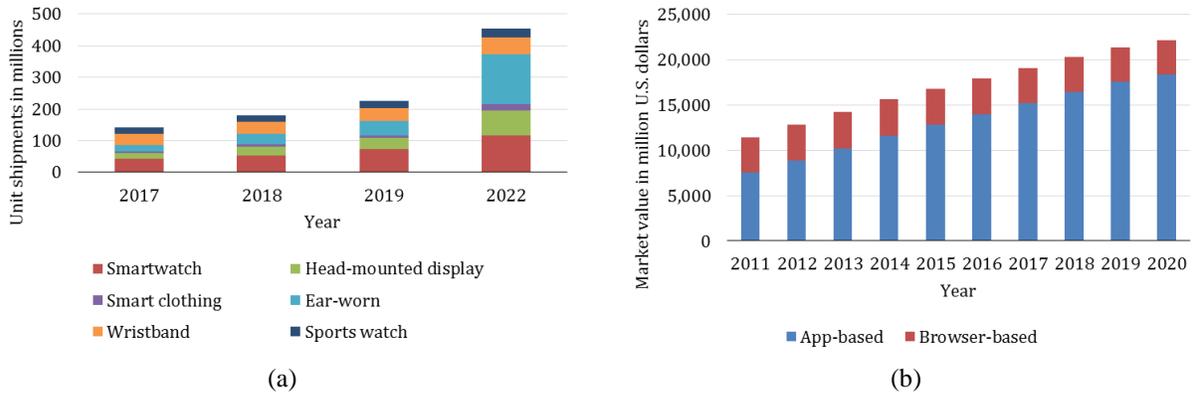

Figure 4. Trends of inter-connectivity: (a) Forecast unit shipments of wearable devices worldwide from 2017 to 2022 [20]. (b) Value of the social gaming industry worldwide from 2011 to 2020 [21].

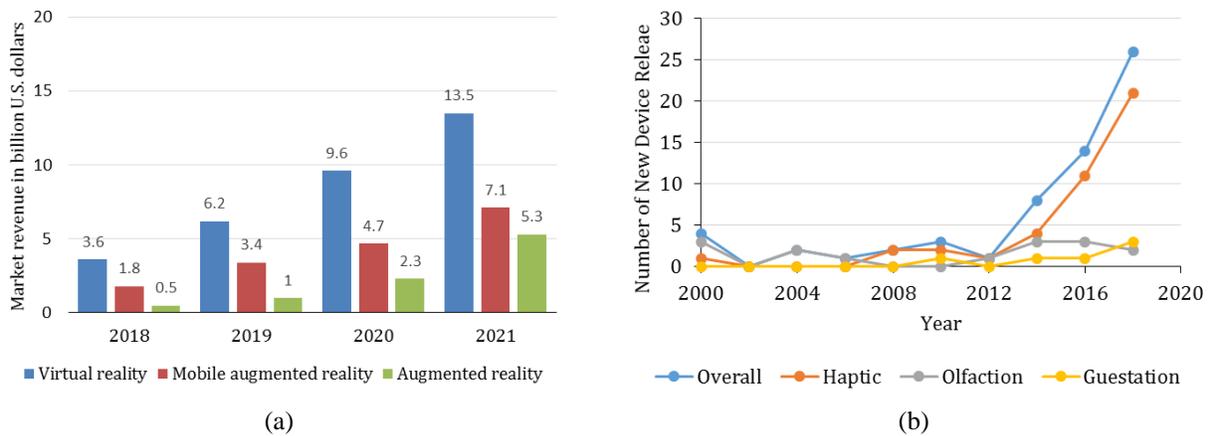

Figure 5. Trends of the immersive technology and market: (a) Immersive technology consumer market revenue worldwide from 2018 to 2021 [22]. (b) Number of new virtual reality sensory device releases along with years, by category. All the figures before 2000 are aggregated to 2000. See the complete list in the supplemental material.

## 2.5. Intelligence

Game intelligence (i.e., Game AI) techniques have been widely applied in virtual worlds for creating the sense of challenging, enriching human-computer interaction, providing better content, and bringing extra experience to users [23]. A previous study has rendered the milestones of game

AI innovation in NPC behaviour, procedural content generation (PCG), and narrativity (in the supplemental material). Game AI in NPC can endow NPCs with believable or human-like characteristics in interaction. The game AI in procedural content generation (PCG) can automatically generate a complex world system, many 3D models, and a full scenario with random quests. Moreover, game AI can also develop game plots from interactive storytelling to augment users' dynamic and immersive gaming experience. To further improve game AI, machine learning techniques have been applied to games. They have become either the dominant or the secondary method in NPC behaviour learning, player modelling, PCG, computational narrative, believable agents, AI-assisted game design, and general game AI [23]. Summerville et al [24] has surveyed the cutting-edge machine learning approaches applied in PCG. Besides, some artificial intelligence techniques have been applied in story generation [25], such as genetic algorithms. The above trends show that we can witness the increasing adoption and improvement of intelligent techniques in virtual worlds.

## 2.6. Implications

The mobility trend implies capabilities of heterogeneous client-end devices' for running virtual world applications. Compared with desktop devices and game consoles, mobile devices are configured with fewer hardware resources. In terms of floating-point arithmetic computation, for example, the performance of Nintendo Switch (2017, portable mode) and Samsung Galaxy S10 (2019) is 157 GFLOPS and 899 GFLOPS respectively, whilst Sony PlayStation 4 has already reached 4197.8 GFLOPS in 2014. On the other hand, the immersion and intelligence trend imply the increase of rendering complexity and computing tasks. The conflict between the limited client-end resource and the rapid increase of computational complexity implies that existing clients may have limited capability to provide highly immersive rendering and intelligent computing. Thus, computing resource sufficiency is the first emergent requirement.

Moreover, mobile devices are more subject to failure than their desktop counterparts for two reasons. First, a connection to remote services through a wireless network is less stable that through a wired network, as the former relies on signal quality. Second, mobile devices have limited battery life and battery quickly depletes for resource-hungry applications, such as virtual world, leading to device failure. A poorly connected or highly congested network will even cause extra power consumption, due to communication overhead increase [26]. Since unstable connection will corrupt users' gaming experiment, and device failure may cause game state lost, computing reliability is another new requirement.

Diversity implies that more content will be generated by users, including user-created virtual objects and virtual wealth generated in gaming [27]. Meanwhile, interconnectivity implies that more user information will be stored in virtual worlds, including the user data collected from multiple peripheral (i.e., wearable and IoT) devices and the relationships built from users' online social network. For simplicity, user-generated objects, virtual wealth, user data, and use social connections are uniformly called user content. The increase and importance of user content have two questions drawing our attention. First, intuitively, will user content be lost if a virtual world application is failed, such as infrastructure failure or service discontinuation? To avoid content lost, this question brings the persistency requirement to system design.

Besides content storage, users may also care more than ever about how a system will treat their content. With the increase of personal data and social connections, data security is gradually becoming a great concern to users, especially the security of privacy-sensitive information, including personal identification, personal (i.e., religion, race, health, sexual orientation, financial, biometric, etc.) information, collected data, and device traceable information. With the increase of virtual properties (i.e., objects and wealth), legal protection has become another concern, especially for those having real-world economic value (e.g., the Linden Dollar in Second Life). In the case of service termination, for example, users may worry about the access to or even the economic benefit from them [28]. Moreover, users may share their content with others, leading another level of security and legal concern. All the above concerns request for a trusted environment of content storage. We call it the credibility requirement.

Bringing together the above implications, Figure 6 summarizes the relations between the emergent technology trends and requirements of virtual world development, including sufficiency, reliability, persistency, and credibility. They will be elaborated in the next sections.

$$Mobility \land (Immersion \lor Intelligence) \rightarrow Sufficency$$
$$Mobility \rightarrow Reliability$$
$$Diversity \lor Interconnectivity \rightarrow \begin{cases} Persistency \\ Credibility \end{cases}$$

Figure 6. Relation of emergent trends and design requirements.

## 3. New Requirement Space

The trend analysis has implied new requirements of virtual world system design, which are complementary to the existing requirements, namely consistency, responsiveness, scalability, and security. We have observed that responsiveness is a common criterion in the sufficiency, persistency, consistency, and scalability problems. Hence, it is more appropriate to express it as a design criterion towards other requirements. Figure 7 shows the taxonomy of design requirements and issues.

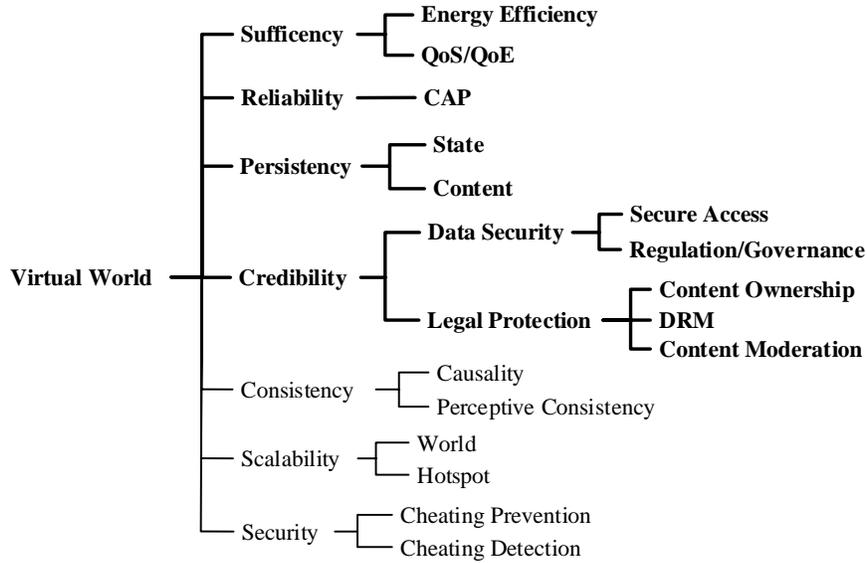

Figure 7. Taxonomy of design requirements and issues.

## 3.1 Brief Introduction on Existing Requirements

Before elaborating on the new requirements, existing requirements are briefly introduced for completeness. The consistency models originate from distributed computing [29]. For virtual world systems, consistency can be studied in two domains [30]. In the discrete domain, it requires that two users can see the same set of events handled in causal order, called causal preservation. Yet, in the continuous domain, object state also changes over time, following some internal rules, e.g., physical laws. It then requires that two users can see the same state changes if they have received the same set of events, called perceptive consistency [31]. In practice, consistency error can be compensated by latency error to achieve the overall fairness, which is called playability [32]. Scalability is "a desirable … ability of a system to accommodate an increasing number of elements or objects, to process growing volumes of work gracefully" [33]. In virtual worlds, the scalability issues can be classified into world scalability and hotspot scalability. The challenge of world scalability comes from two sources [34]: large scale real-time simulations and simultaneous visualization for many users, caused by user population increase. Hotspot scalability comes from another source: multiple heterogeneous actors with different operating characteristics. When many users gather at a small place, a hotspot is formed [35]. The scalability issues aim at minimizing system performance reduction with the increase of users and user interactions, either globally or regionally. For security, cheating is an ubiquitous issue in online gaming communities [36], caused by the combination of malicious attempts and system faults. [37] has classified them into 3 types of security break and further divided them into 13 categories. They summarized that cheatings in virtual worlds need either be prevented or be detected for providing a fair playground.

## 3.2. Sufficiency

Sufficiency is the first emergent requirement such that users can access highly immersive environments from any device. It becomes outstanding with the growth of mobile devices and computational complexity. Partially/completely offloading computing tasks to remote sites for meeting the resource requirement is an evident choice [38], [39]. Based on the choice, two concerns can be derived: energy efficiency and quality of service (QoS)/quality of Experience (QoE). Energy efficiency concerns about the device energy consumption for local computing or communication to remote sites [40]. [41] provided an energy consumption model for smartphone task offloading in WLAN and mobile network (3G/4G). QoS/QoE concerns about the cost and benefit of offloading [39]. The cost includes communication overhead and response latency. The benefit includes graphic (i.e., image or video) quality improvement, quantified by bitrate or frame rate. Moreover, both energy efficiency and QoS/QoE are sensitive to network condition. If network availability is low, code offloading will increase both energy consumption and latency [26].

### 3.3. Reliability

Reliability also comes from the mobility trend, as both mobile devices and mobile networks are subject to failure. Redundancy can add reliability to a system by replicating data and program to multiple sites so that a device can both tolerate connection failure (i.e., failure tolerance) and recover application state after client failure (i.e., failure recovery). Redundancy brings the consistency problem (called replica consistency to distinguish it from the consistency requirement), since all replicas must maintain the application game state for state integrity. Consistency is a hard problem in distributed computing, due to the notorious CAP theorem [42]. It states that if a design cannot tolerate network partition, availability and consistency cannot be simultaneously achieved, because a network failure may prevent a replica from being synchronized [43]. On the other hand, if a design can solve the network partition problem, consistency and responsiveness will then become the conflicting requirements, since consistency control protocols normally add additional communication steps [44]. Thus, the reliability property seeks for a computing redundancy design which can achieve high availability, consistency, and responsiveness, meanwhile without or tolerating network partition.

Additionally, reliability is an important requirement for device interoperability [45] in pervasive games. In Hot Potato [46], for example, player mobile devices are locally connected with wireless sensor network (WSN) and P2P neighbour discovery. They also need to connect to a backbone network for coordinating heterogeneous devices, providing global interaction and storage for all game instances.

### 3.4. Persistency

The persistency requirement has two aspects: state and content. State persistency requires that when a player leaves a game, his/her game state is well-kept for the next-time retrieval and play [47]. State data normally has a small size but could be frequently updated. Thus, data read/write efficiency is the main concern. [48] found that some states only require approximate consistency between write and read (e.g., avatar position), while others require exact consistency (e.g., user inventory). Thus, different persistency strategies can be applied for different consistency requirements to minimize system overhead and response delay. Game state storage also needs to be reliable and robust to any failure. If redundancy is applied for fault-tolerance, then the replica consistency, responsiveness, and load balance issue also have to be studied [49].

On the other side, user content persistency in virtual world is still short of attention. With the growth of user content, however, this aspect is becoming increasingly important. Content persistency requires that a user's data and content can be permanently preserved if the user has not departed from the virtual world. Compared to state data, user content files normally have a larger size (e.g., multimedia files), but are less frequently updated. Thus, the main concern is distinctive [50], which includes storage efficiency (i.e., reliability to storage space), bandwidth cost, data access latency, and content integrity. Furthermore, if user content can be shared to other users, view consistency is needed, which requires that two users can see the same set of objects if both are interested in them.

**3.5. Credibility**

Credibility (or trust) comes from the increase and importance of user content on virtual worlds. It is more complex and thus deserves some detailed explication. In the context of information system, trust is a subjective belief that a system or a system component will behave as expected [51], which is established based on direct evidence, recommendation, and reputation [52]. In virtual world, users can have their belief on the underlying system in content management only if their content is securely preserved and legally protected. Thus, the credibility requirement has two concerns: user data security and content legal protection.

Data security (to distinguish it from the security requirement), though new to virtual worlds, has been widely studied in cloud services, which includes privacy and confidentiality. Both require that user's sensitive content is confidential to unauthorized access, whilst data privacy stresses more on the interest of users in controlling the data about themselves [53]. [54] divides the top threats into 5 categories, including data breaches, account or service traffic hijacking, API or shared technology vulnerabilities, malicious insiders, and insufficient due diligence. The categories of threats imply that data security is not only a technical issue but also a regulatory issue. From the technical perspective, secure data access needs to be achieved by mitigating the above threats from privacy-enhancing technologies, security mechanisms, data obfuscation, or anonymization. From the regulatory perspective, data security involves procedural solutions for committing legal and contractual requirements in a verifiable way. For verifying requirement commitment, accountability [55] and compliance are the key issues.

Legal protection concerns the legitimate interest of the involved parties, mainly virtual world users and platforms. The first issue is content ownership. If users only have the control of their created content but not the ownership, such separation will become the barrier to user innovativeness due to the additional cost [56]. Secondly, if users have content ownership, they may need to control the distribution of their content to others through digital rights management (DRM), which only allows users to represent and constrain content usage but also provides traceability of service violation. Moreover, user-created content may also bring legal risks, including plagiarism, offensive content, spam, soft hacking, etiquette breach, personal exposure, etc. [57]. These risks push the content moderation techniques to a new frontier.

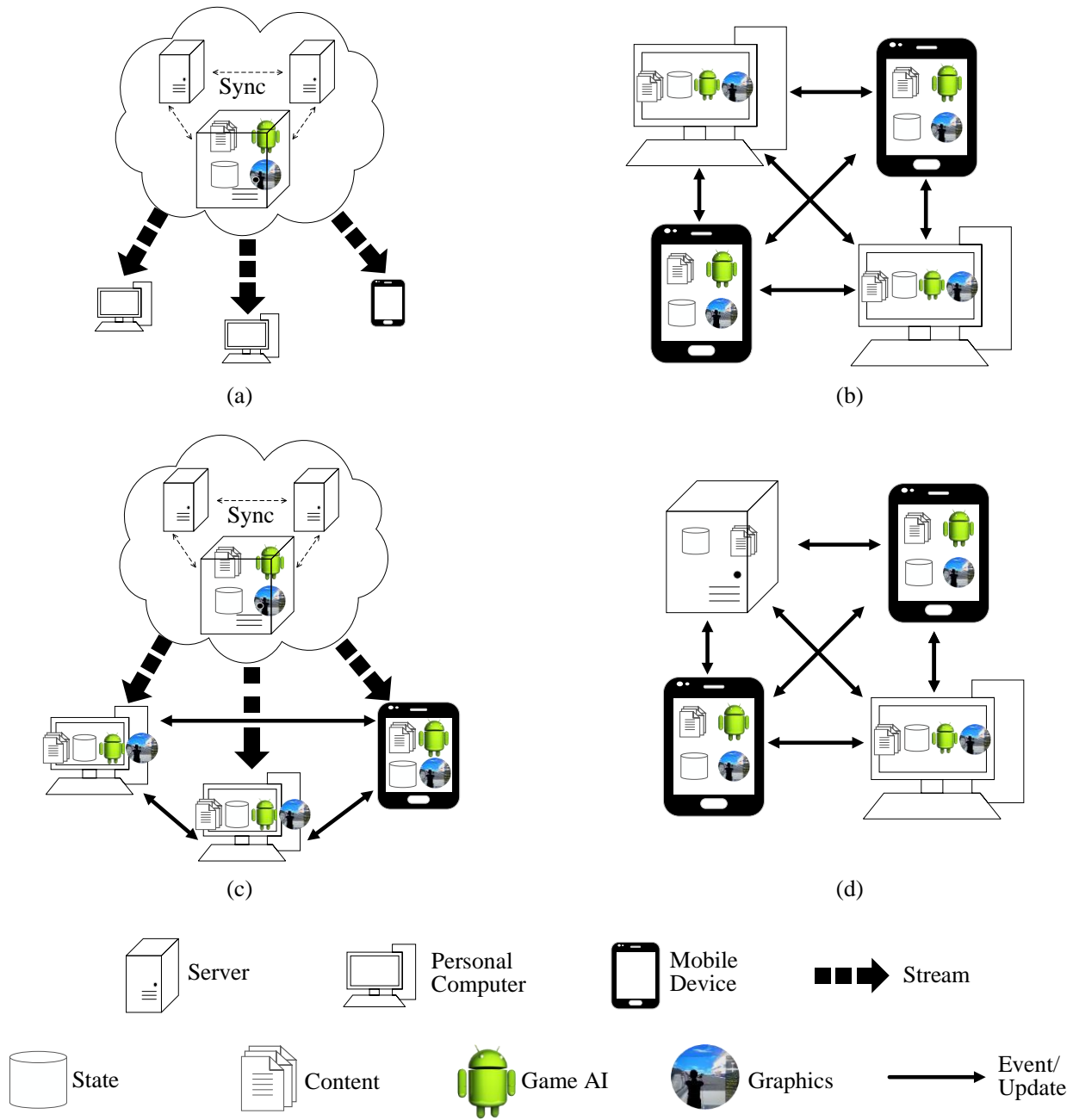

Figure 8. Conceptual classes of virtual world architecture: (a) client-server, (b) peer-to-peer, (c) Hybrid-I, (d) Hybrid-II.

## 4. Virtual World Architecture Inspection

Existing virtual world system architectures can be classified to client-server (C/S) architecture, peer-to-peer (P2P) architecture, and hybrid architectures. Figure 8 illustrates the conceptual structures of them. This section introduces the existing virtual world architectures and conducts a qualitative study on their design requirement satisfaction and limitations. Table 1 lists the representative solutions to the design issues. In this paper, only the issues related to emergent requirements are elaborated.

Table 1. Representative solution of the design issues.

| Requirement | Design Issue | C/S | Hybrid-I | P2P | Hybrid-II |
|---|---|---|---|---|---|
| Sufficiency | Energy Efficiency | Energy trade-off [40] | | | |
| | QoS/QoE | MAUI [38] | | | |
| Reliability | Reliability | Database replication [58] | Database replication [58] | Pithos [59] | Backup virtual node [60] |
| Persistency | State Persistency | | Centralized control [61] | | State Action Manager [60] |
| | Content Persistency | | | Kad [62] | |
| Credibility | Secure Access | Policy-based access control [63] | | Blockchain-based access control [64] | |
| | Regulation | Data compliance [65] | | N.A. | |
| | Ownership | User representative [66] | | | |
| | DRM | License Server [67] | | DRMChain [68] | |
| | Content Moderation | Content approval or social accountability [57] | | | |
| Consistency | Causality | Local-lag and timewarp [30] | | | |
| | Perceptive Consistency | | | | |
| Scalability | World Scalability | Cloud gaming [69] | Cloudlet [70] | Red-Black [71] | Cloud resources [60] |
| | Hotspot Scalability | KIWANO [72] | Peer-assisted [61] | AOI-cast [73] | Virtual server transfer [60] |
| Security | Cheating Prevention | Cryptographic measures [74] | | | |
| | Cheating Detection | Behaviour monitoring [75] | IRS [76] | Mutual verification [77] | RACS [78] |

## 4.1. Client-server

In the C/S architecture (Figure 8 (a)), simulation logic is processed at the server end, while clients only render updates, which are sent from servers, to users. A C/S virtual world has a definitive platform owner (e.g., Linden Lab, the owner of Second Life). The online services (e.g., authentication, inventory management, and region management in Second Life) and application data (e.g., user items, user inventory, and virtual property in Second Life) belong to the platform owner. C/S architecture has predominated the virtual world industry (See the list of sampled virtual worlds in the supplemental material).

**Requirement satisfaction**

C/S architecture adopts a centralized structure, which easily satisfies sufficiency and reliability. For sufficiency, codes running on client can be partly or completely offloaded to dedicated servers through function invocation[79]. In code offloading, [40] has provided an energy trade-off model to maximize energy efficiency in terms of the ratio of energy cost to device-specific cost in local computation and additional communication. For QoS/QoE, MAUI [38] has shown the improved video refresh rate from 6 FPS to 13 with low energy consumption, latency, and data transfer overhead. Cloud gaming [39] can also address the sufficiency issue through interactive remote rendering [80]. As such, energy consumption and QoS/QoE remain the main focuses of cloud gaming.

C/S virtual worlds can resolve the reliability problem with replication. [58] applied a database replication approach to multiplayer online games to provide fault tolerance. Game operations are handled with a Read-One-Write-All-Available approach. That work provides four alternative synchronization approaches for propagating updates to all replicas to maintain consistency. It also shows fast responsiveness (specifically, ≤20 ms for read-only transactions and ≤100 ms for write transaction). Database replication can also improve reliability in game state storage. Yet, to the best of our knowledge, existing C/S virtual world systems have not fully addressed the connection failure issue. Directing communications to nearby fog facilities might be a solution [81], since a shorter distance between fog facilities and mobile clients can reduce the chance of connection failure.

For data security, policy-based access control [63], together with secure hardware [82] or cryptographic measures [83], can be applied in data sharing [83], keyword search [84], and public audit [85]. Fog computing can also mitigate insider data theft [86]. Moreover, security policies can be enforced through data protection compliance [65] and accountability check [55]. For legal protection, Kim et al [67] proposed a user-centric rights representation model for the DRM of user-generated content by employing a separate license server [67].

Moreover, the content moderation issue has been discussed in [57], which proposed a risk management framework for user-generated content with 7 moderation techniques, respectively employed by platform owner, system, trusted curator, and the public. Notably, the solution to this issue is applicable to all architectures.

**Limitations**

The centralized ownership creates a single point of failure, leading to two issues. The first issue is persistency. When the entity owning a virtual world dies, bankrupts, or withdraws its operations, the affected virtual worlds will collapse together with the loss of its user-generated contents (Among the 126 virtual world applications based on OpenSimulator[2], 28 of them has no longer in operation). The result is that user content will become non-persistent. Temporary infrastructure failure may also cause user content lost. For example, the World of Warcraft game server crash on 21 Jan 2006 led to the inventory damage of thousands of players [58].

The second issue is credibility, which involves information and legal protection. User data of a virtual world may include sensitive information, such as location information in context-aware games [5]. One of the key privacy compliance issues is to provide a transparent and controlled environment to data owners [87], which is, however, at the cost of platform owner, and subject to many external factors (e.g., global societal trends), firm-specific factors (e.g., industry best practice and corporate culture), and ethical frame (e.g., rule utilitarianism or moral relativism) selection [88]. There are also some trust-level issues related to transparency, including requirement propagation along the sub-contract chain and malicious insiders [87]. Thus, the gap remains between platform owner's claims and user's trust of their sensitive information protection.

---

[2] See the inactive virtual world list in http://opensimulator.org/wiki/Grid_List

Moreover, storing user-created content on a C/S platform separates content ownership and control in social worlds, creating a legal dilemma. Users do not have the control over the actual data and a platform may unilaterally terminate their accounts with their virtual asset confiscated, as in the Bragg case [89] and the Evans case [90], which leads to ownership tensions and inhibits user innovation [66]. [28] have shown that platforms tend to have such ownership separation either for competition purpose or for liability safe harbour, which discloses the second problem. Storing user content imposes legal pressure on platform owners for copyright infringement or offensive content [91]. Though content moderation can mitigate such risks [57], it largely increases the cost paid by platform owners and the risks still exist.

### 4.2. Peer-to-peer

P2P virtual world aims at resolving the scalability issue of the classic C/S architecture [92], which is jointly promoted by P2P computing and virtual world researches. In the P2P architecture (Figure 8 (b)), services are collectively run by user devices which play the role of both server and client. All user devices are connected to a P2P overlay network [93]. When a user is accessing a virtual world application, he/she is also providing services to others of the same application through his/her accessing device. Such a virtual world is supported by a reciprocal economy. A P2P virtual world can have an application provider who develops and distributes the application software to run on a P2P overlay network. Currently, P2P virtual worlds are still far from wide industry acceptance, partially due to their higher difficulty on system maintenance for developers [37].

**Requirement Satisfaction**

P2P virtual worlds, firstly proposed in Knutsson et al [92], remove the single point of failure of C/S virtual worlds. By running applications on user devices with open protocols, application data are no longer owned or controlled by any entity. Thus, they are promising to address the persistency issues. [49] provided a comprehensive survey of P2P game state persistency. Later, [59] proposed a two-tier architecture. An overlay storage employs a distributed hash table (DHT) and super-peers to provide high reliability and fault-tolerance. A group storage employs a distance-based grouping approach to improve availability and responsiveness.

For content persistency, P2P virtual worlds can also employ a P2P file storage protocol to store user content. Existing P2P file storage systems, such as BitTorrent (bittorrent.com), has shown their persistency property without central storage. Varvello et al [62] designed a communication infrastructure for distributing and storing virtual objects to Kad, a global P2P storage network. Total Recall [94] provides optimized storage efficiency with high responsiveness and low bandwidth cost. [95] proposed a content retrieval approach based on Total Recall to provide efficient content integrity check. These properties are requested by the content persistency requirement. Moreover, [96] proposed a P2P 3D object streaming scheme from nearby peers, which can be used in virtual world content sharing. [97] improved it in efficiency.

For data security, P2P virtual worlds can leverage Blockchain technology, combined with cryptographic measures, to achieve user-centric secure data access [98]. Blockchain-based privacy-preserving access control has been widely discussed in many fields, including personal identity management [99], healthcare record sharing [64], collected data access [100], etc. Blockchain technology has some advantages. First, a third-party service is not needed for stewarding user data, on which users must put their trust. Second, security policies added to Blockchain [101] is transparent to all users and data compliance can be achieved through the consensus mechanism. Moreover, a Blockchain data structure is tamper-proof, providing additional merits, including integrity and non-repudiation, etc.

Blockchain technology can also be used in digital rights management [68]. Its non-repudiation property can enable the conditional tractability of license violation. Moreover, a P2P virtual world does not have a definitive platform owner, attributed to its decentralized structure. It removes ownership inconsistency issue and users really have the ownership of their virtual assets as well as the underlying data. Thus, a P2P virtual world can achieve higher credibility.

**Limitations**

Sufficiency and reliability are the two weaknesses of P2P architecture, and they are becoming outstanding with the increase of mobile clients. In pervasive games or augmented reality, user clients could be wearable devices, IoT devices, and custom-built devices. They have nonstandard interfaces and heterogeneous capabilities [5], [45], which exhibits several limitations in the P2P design. First, resource-limited wearable devices may not have enough computing resources or standard interfaces to play the role of a server for running full game logic. Yet, gaming experience on mobile devices can be unsatisfying because of limited processing capacity. For example, Final Fantasy XV Pocket Edition provided lower graphics [102] and AI experience [103] than the desktop and console edition.

Moreover, P2P virtual worlds may also suffer from heterogeneous peer resources. Mobile devices and wearable devices can run fewer services than desktop devices. Departure of users with major resources provision can unduly stress the rest peers [104].

Mobility also greatly increases the chance of client failure due to connection lost [105] or battery depletion [106]. Client failure may cause unsaved state loss. Also, reliability is important to gaming experience. Pervasive games require high connectivity to game masters (i.e., specially selected players) for content distribution, diegetic communication, and game progress track [45]. Although Pithos [59] adds reliability through a two-tier game state replication, the design does not address the replica synchronization issue, which may lead to inconsistent game state.

**4.3 Hybrid-I**

The C/S and P2P architecture can be combined to a hybrid architecture to overcome the weakness of one architecture by the strength of another. According to the ways of combination, hybrid architectures can be divided into two classes. The first one, denoted by Hybrid-I (Figure 8 (c)), leverages the P2P computing techniques to the C/S architecture for computation offloading, since P2P resources can be easily scaled with user population. In Hybrid-I virtual worlds, e.g., [61], clients disseminate updates to each other through P2P communication to save a server's outgoing bandwidth and thus its operating cost. Cloud gaming can also exploit P2P techniques to reduce game latency [39] and service operating cost [70], [107].

**Requirement satisfaction and limitations**

The Hybrid-I architectures have different requirement satisfactions, depending on how clients participate in a simulation. This section introduces some typical examples. Firstly, in [108], the cloud servers form a P2P publish-subscribe overlay for the communication between game servers. Clients do not involve in any simulation. Then, in [61], a central server is employed to control and store game state, while clients are only in charge of message dissemination to save the server's outgoing bandwidth. Their requirement satisfaction is similar to that of C/S architecture, except that the latter design provides more cost-efficient scalability solutions, because the efficiency of message dissemination scales with user population. They also share the same single point of failure with C/S architecture due to the existence of a central point for content storage, including the limitations in content persistency, data security, and legal protection.

In [76], a central server stores application states and relays client messages for security check. Clients compute and disseminate the game state with P2P techniques. The requirement satisfaction of this approach lies between the C/S architecture and the P2P architecture. Specifically, the P2P clients allow it to have similar requirement satisfaction and limitation to the P2P architecture. The centrally controlled game state allows it to have the similar requirement satisfaction and limitations to the C/S architecture in persistency, credibility, and security, while P2P clients allow it to have the similar requirement satisfaction and limitations to the P2P architecture with respect to the rest issues.

In [109], moreover , P2P clients manage application state for each other, while a central server only keeps the sensitivity data (e.g., user profile) and provides the utility functions (e.g., authentication). This design is similar to the P2P architecture, except that the satisfaction of the data security requirement can approach to that of the C/S architecture because of the centralized storage of sensitive data.

### 4.4. Hybrid-II

The second hybrid class, denoted by Hybrid-II (Figure 8 (d)), introduces cloud resources to the P2P architecture to improve the overall performance. In [110], for instance, special server nodes play the roles of zone masters. They do not retain game state but only help clients in intra-/inter-zone communication. The Hybrid-II architecture was initially proposed by [104] after identifying the heterogeneous peer resource issue. In their design, cloud resources are introduced to complement peer resources, and both are virtualized to nodes (called virtual nodes, VN). The VNs construct two P2P overlays for providing the common services: the state action manager (SAM) and the positional action manager (PAM). The SAM nodes, organized to a structured overlay, manage the state of virtual objects [60]. The PAM nodes, organized to an unstructured overlay, manage user position for neighbour and object discovery [111]. In an application of Hybrid-II architecture, a client firstly receives the nearby users and objects by querying the PAM service. Then the client can interact with them through the SAM service for state update and state synchronization.

**Requirement satisfaction**

The Hybrid-II architecture shares some similarity with the P2P architecture, but there are some differences between them. Firstly, the Hybrid-II design has enhanced the scalability by resolving the resource heterogeneity issue with reliable cloud resources to cover resource deficiency from user departure. The optimal cloud resource assignment problem with respect to system load has been studied in [112] to minimize the economic cost from cloud resource utilization. Resource virtualization can also provide load-balance to further improve scalability. In [60], load imbalance is evaluated with the Gini Coefficient. If a device is overloaded, it will migrate some of the VNs to other devices with minimized visual inconsistency.

Cloud resources also have a contribution to state persistency [104]. The SAM service provides two layers of replication. First, each VN assigned to a user device (called uVN) is backed up with a cloud node (called bVN). An uVN periodically synchronize its node state to the bVN. Such fault tolerant mechanism is called coarse-grain data replication. Moreover, a VN in SAM also dynamically replicates its objects to a VN (i.e., overlay replicas) in the neighbour address space [113], which is called fine-grained data replication. In the case of uVN failure, the SAM service will forward requests to the overlay replica. If an overlay replica has become overloaded, it will inform its clients to forward their requests to the bVN. In such a design, the coarse-grain data replication can reinforce storage reliability, and the fine-grain one can improve service responsiveness through neighbour address access [113].

Though not mentioned by the authors, we believe that content persistency can be achieved with the same approach as that in the P2P architecture, since the overall Hybrid-II architecture is decentralized. Likewise, the Hybrid-II architecture can achieve the same credibility requirement satisfaction as the P2P architecture, due to the lack of a centralized control entity.

**Limitations**

Similar to the P2P architecture, the Hybrid-II architecture is limited in resource provisioning for resource-limited devices. In the SAM service, objects stored on a VN have the identifiers close to the VN's address [60]. That is to say, objects are not group by their users but by mapping identifiers to the address space of the DHT overlay [113]. Thus, a SAM node may not have all the content of a user for running complete simulation logic. Moreover, even though cloud resources have been introduced, which may have higher computation capabilities, they are partitioned to many VNs for service and data storage backup. Thus, cloud gaming cannot be run on the Hybrid-II architecture for code offloading [79] or interactive remote rendering [39].

For reliability, though the two-layer data replication mechanism can double the reliability of the SAM service, as in the P2P architecture, the state synchronization issue between replicas is not addressed by the authors. When an uVN fails, it may not timely and successfully synchronize the latest state to its bVN or overlay neighbour, leading to state lost or inconsistency. The extent of inconsistency between replicas depends on the length of synchronization interval. However, state consistency could be critical to the integrity of a game, such as in-game trade and user inventory [48].

## 5. Discussion

### 5.1. Architecture Comparison

In this section, we compare the virtual world architectures in requirement satisfaction. The introduced architecture classes include C/S, P2P, Hybrid-I, and Hybrid-II, and Virtual Net. In the discussion, they are further classified to centralized architectures (C/S and Hybrid-I) and the decentralized architectures (P2P and Hybrid-II). Table 2 shows the comparison result.

Table 2. Features comparison of virtual world architectures.

| Requirement | Feature | C/S | P2P | Hybrid-I | Hybrid-II |
|---|---|---|---|---|---|
| Sufficiency | Lightweight Device Support | Code offload, Cloud gaming | No | Code offload, Cloud gaming | No |
| Reliability | Connect Failure Tolerance | No | No | No | No |
| | Replica consistency | Yes | No guarantee | Yes | No guarantee |
| Content persistency | Content lost due to platform failure | Likely | Unlikely | Likely | Unlikely |
| Credibility | Transparency to Compliance | No | Yes | No | Yes |
| | Content Ownership | Separated | User-owned | Separated | User-owned |
| | Platform Owner's Liability for illegal content | Yes | No | Yes | No |
| | DRM Violation Traceability | No | Yes | No | Yes |
| Scalability | Scalability Cost to Application Provider | High | Low | Moderate | Medium to low |
| Security | Client-side Code Tampering | Highly secure | Vulnerable | Highly secure | Vulnerable |

Sufficiency can be satisfied with the separation of client and service. In centralized architectures, since most computational complex tasks are moved to the service end, including simulation logic, physics computation, and graphics computation, a lightweight client only needs to render the results generated from servers. In contrast, the decentralized architectures require client devices to play the role of both server and client, imposing large computation load on them. A lightweight client device may not have the capability to render a highly immersive environment. Moreover, compared to the rest architectures, P2P virtual worlds may suffer from heterogeneous peer resources. Some peers may be located on low-performance or unreliable devices, which will provide less services to other peers and even slow down the entire system.

Reliability can be satisfied with replication and synchronization. In centralized architectures, each state update will be synchronized from client to server. Since the centralized architectures preserve a copy of user state at the server end, they can tolerate client failure. A recovered client can catch up with the latest state by retrieving the data from server. Moreover, the database synchronization approaches can maintain replica consistency between replicated databases. But centralized architectures only maintain a connection between a client and a server, which does not tolerate connection failure. The decentralized architectures also provide replication for backing up client state. Thus, client state can be recovered after client failure. But they neither provide replica consistency guarantee nor tolerate connection failure.

Content persistency can be satisfied with decentralization. The centralized architectures contain a central entity for stewarding user content, creating a single point of failure. As above-shown, even if advanced fault-tolerance techniques can minimize the failure of a system, the failure of the control entity will cause services to discontinue. Decentralized architectures, on the other hand, are immune from system-level failure, since such a system is not controlled by a single entity. Moreover, the underlying P2P storage techniques [94] ensure that the failure of a node will not spread over the entire system, and local failure can be recovered from data replicas.

Data Security can be satisfied also from the decentralization. The security compliance of the centralized architectures is normally non-transparent to the external and is subjective to many firm-specific factors. Users only put limited trust in the protection of their data and information, even for cloud-based systems. An industry observation has shown that security and compliance are still the significant challenges to cloud computing in 2019 [114]. The decentralized architectures, on the other hand, does not have a single stakeholder. Every user is the stakeholder of a system, and P2P-based security mechanisms can be applied and effective as long as most users are obedient to the system rules [115]. Thus, users need to trust no entity but only the system per se, which is also the argument why Blockchain-based solutions can achieve higher security in data access [64]. Thus, the decentralized architectures can provide more trust in data security to users.

Legal protection can be satisfied with unitary ownership. In the existing virtual worlds, which are centralized, control entities own all digital assets (i.e., file and data), while users own their virtual assets, leading to ownership inconsistency and disputation [66]. The decentralized architectures, on the other hand, do not have a control entity. Both virtual objects and digital assets belong to users. Thus, users' virtual property can be legally protected by property or copyright law. Moreover, application providers do not need to worry about their liability for, e.g., plagiarism or offensive content, since they do not own them. Thus, the decentralized architectures can offer higher legal protection to both users and application providers.

Table 2 also shows the architectural comparison for existing requirements, including scalability and security. Scalability cost refers to the economic cost imposed on someone or some entity, who provides the virtual world application, for scaling up the system to accommodate the growth of the user population. In the C/S architecture, the cost is solely paid by the platform owner who pays the cost of the entire infrastructure. In the hybrid-I architecture, since some functions (e.g., state distribution) are distributed to user clients, the platform owner's cost can be largely decreased. In the hybrid-II architecture, the cost depends on who provides cloud resources. Since a hybrid-II virtual world is collectively run by cloud resources and user devices, the cost in the hybrid-II architecture will not be larger than the one in the hybrid-I architecture. In the P2P architecture, the infrastructure cost is fully shared by all users. Thus, application providers barely need to pay for the infrastructure.

Game cheating is the main factor leading to the low acceptance of decentralized architectures in the industry [37]. Compared to centralized structures, decentralized structures, due to the lack of a central arbitrator, are more vulnerable to player escaping, network flooding, suppress-correct cheat, etc. than their C/S counterparts, while time cheat, blind opponent, consistency cheat are only possible to P2P games [37]. In P2P architectures, though some cheatings can either be prevented or detected through cryptographic measures [74] or mutual verification [77], a more general approach either needs a central server for rule enforcement [76], [116] or lockstep message check which largely increasing communication overhead. In the Hybrid-II architecture, a referee anti-cheating scheme is applied for detecting illegal messages in communication [78]. Nevertheless, a malicious user still can tamper the code of VNs and change the simulation logic which becomes favourable to them and cannot be detected by the scheme. Moreover, the deterministic mapping from object identifiers to the VN address space [113] enables an attacker to guess the content on the controlled device, increasing the attack success rate.

## 5.2. A Possible Avenue Towards Future

For future virtual worlds, a new architecture called Virtual Net is under development to satisfy more requirements[95]. The central idea of Virtual Net is that nobody owns a virtual world, but all users collectively create a self-organized one, which is like the decentralized architectures at this point. In Virtual Net, users contribute a part of their computing resources, which is virtualized into one or multiple virtual nodes. All virtual nodes have the same computing resources, managed by a node pool. Users of Virtual Net can store their contents or deploy their applications on the nodes without a central server. Thus, it can inherit the advantages from the decentralized design paradigm in persistency, data security, and legal protection.

In the Virtual Net architecture, a Mesh is a set of replicated virtual nodes. One Mesh is assigned to each user for running virtual world applications. The replicas in a Mesh applies a replica synchronization protocol to maintain replica consistency. Thus, externally, each Mesh can be treated as a reliable peer, and inter-Mesh interaction is equivalent to the inter-peer interaction in the P2P architecture. In a Mesh, a client is a special node, which provide the user interface for receiving user operations and rendering updated states to users. A Mesh can receive the operations from or send the updates to a client. Such structure offloads the computational complex tasks from the client to the remote service end. So, a lightweight client device only needs to concentrate on limited computing tasks. Moreover, a client can simultaneously communicate with all the replicas of a Mesh to provide connection failure tolerance. Thus, the overall reliability of Virtual Net is higher than the existing architectures. Notably, a special Mesh is be assigned to an NPC for running AI program, which does not have a client node. Virtual Net also contains a P2P cloud which provides the common services needed by all Meshes, such as object request routing [117].

Though promising, a full-fledged Virtual Net requires various challenges to be resolved, including virtual resource management, rapid replica synchronization, composition of event handling and object simulation. Currently, the Virtual Net research is making progress [118] and we can anticipate a better future of virtual worlds.

# 6. Conclusions

We have analysed the trends of the virtual world related technologies and their implications for future virtual worlds. The importance of trend analysis lies in the proposal of new requirements that have not been commonly recognized in the virtual world community. Based on the trends and implications, we have discussed the emergent requirements of virtual world system design in detail. These requirements, including the design issues and criteria, provide a complete requirement space for design reference. We have also thoroughly examined the existing virtual world architectures and discussed their satisfiability and limitations to the new requirements. A complete list of requirement satisfaction examples has been provided for function implementation reference. Then, the comparison between architectures shows that none of the existing architectures can fully satisfy for all requirements. The detailed comparison results provide new avenues for virtual world system development. We hope that the results of this study, the surveyed content, and the analysis can lay a solid foundation and research avenues for future virtual world development.

## Acknowledgments

This research is partially supported by the University of Macau Research Grant No. MYRG2017-00091-FST.

## Supplementary Materials

The source data collection can be retrieved from https://sunniel.github.io/VirtualNet/post/trend-analysis/Supplemental_Material.pdf

## References

[1] J. D. N. Dionisio, W. G. B. III, and R. Gilbert, "3D virtual worlds and the metaverse: Current status and future possibilities," *ACM Comput. Surv.*, vol. 45, no. 3, pp. 1–38, Jul. 2013, doi: 10.1145/2480741.2480751.

[2] G. Fitzpatrick, S. Kaplan, and T. Mansfield, "Physical Spaces , Virtual Places and Social Worlds : A study work in the virtual," in *CSCW '96 Proceedings of the 1996 ACM conference on Computer supported cooperative work*, 1996, vol. 96, pp. 334–343, doi: 10.1145/240080.240322.

[3] K. J. L. Nevelsteen, "Virtual world, defined from a technological perspective and applied to video games, mixed reality, and the Metaverse," *Comput. Animat. Virtual Worlds*, vol. 29, no. 1, p. e1752, 2018, doi: 10.1002/cav.1752.

[4] C. Girvan, "What is a virtual world? Definition and classification," *Educ. Technol. Res. Dev.*, vol. 66, no. 5, pp. 1087–1100, 2018, doi: 10.1007/s11423-018-9577-y.

[5] V. Kasapakis and D. Gavalas, "Pervasive gaming: Status, trends and design principles," *J. Netw. Comput. Appl.*, vol. 55, pp. 213–236, 2015, doi: https://doi.org/10.1016/j.jnca.2015.05.009.

[6] Statista, "Which of the following devices do you use to play games?," *Statista - The Statistics Portal*. https://www.statista.com/forecasts/790476/devices-used-to-play-video-

games-in-the-us (accessed Mar. 14, 2019).

[7]   Statista, "Share of virtual reality and augmented reality (VR and AR) users in the United States as of 2018, by type of device," *Statista - The Statistics Portal*. https://www.statista.com/statistics/830508/us-virtual-augmented-reality-users-by-device/ (accessed Feb. 15, 2019).

[8]   EEDAR, "Number of mobile gamers in the United States and Canada from 2014 to 2017," *Statista - The Statistics Portal*. https://www.statista.com/statistics/454381/mobile-gamers-number-north-america-device/ (accessed Mar. 14, 2019).

[9]   SuperData Research, "Facebook's social gaming market revenue from 2010 to 2014 (in billion U.S. dollars)," *Statista - The Statistics Portal*. https://www.statista.com/statistics/276454/worldwide-revenue-of-facebook-social-gaming-forecast/ (accessed Mar. 14, 2019).

[10]  M. Hakonen and P. Bosch-Sijtsema, "Virtual Worlds Enabling Distributed Collaboration," *J. Virtual Worlds Res.*, vol. 7, no. 3, 2014, doi: 10.4101/jvwr.v7i3.6158.

[11]  P. Mildner, B. John, A. Moch, and W. Effelsberg, "Creation of Custom-made Serious Games with User-generated Learning Content," in *Proceedings of the 13th Annual Workshop on Network and Systems Support for Games*, 2014, pp. 17:1--17:6, [Online]. Available: http://dl.acm.org/citation.cfm?id=2755535.2755556.

[12]  L. Li *et al.*, "Application of virtual reality technology in clinical medicine.," *Am. J. Transl. Res.*, vol. 9, no. 9, pp. 3867–3880, 2017, [Online]. Available: http://www.ncbi.nlm.nih.gov/pubmed/28979666%0Ahttp://www.pubmedcentral.nih.gov/articlerender.fcgi?artid=PMC5622235.

[13]  D. A. Guttentag, "Virtual reality: Applications and implications for tourism," *Tour. Manag.*, vol. 31, no. 5, pp. 637–651, 2010, doi: 10.1016/j.tourman.2009.07.003.

[14]  T. Kohler, K. Matzler, and J. Füller, "Avatar-based innovation: Using virtual worlds for real-world innovation," *Technovation*, vol. 29, no. 6, pp. 395–407, 2009, doi: https://doi.org/10.1016/j.technovation.2008.11.004.

[15]  Perkins Coie, "Virtual/augmented reality technology and content investment focuses worldwide in 2016 and 2018," *Statista - The Statistics Portal*. https://www.statista.com/statistics/248658/worldwide-mobile-social-gaming-revenue/ (accessed Mar. 14, 2019).

[16]  Eye Create Worlds LLC, "VR for IoT - Rethinking data visualization," 2016. http://www.eyecreateworlds.com/index.php/56-vr-for-iot-and-pushing-the-limits-of-the-dk2-tracking-volume (accessed Mar. 14, 2019).

[17]  Entertainment Software Association, "Essential Facts: About the computer and video game industry," *2016 Sales, Demographic and Usage Data*, 2016. http://essentialfacts.theesa.com/Essential-Facts-2016.pdf (accessed Mar. 14, 2019).

[18]  R. Skarbez, F. P. Brooks, Jr., and M. C. Whitton, "A Survey of Presence and Related Concepts," *ACM Comput. Surv.*, vol. 50, no. 6, pp. 1–39, 2017, doi: 10.1145/3134301.

[19]  A. Covaci, L. Zou, I. Tal, G.-M. Muntean, and G. Ghinea, "Is Multimedia Multisensorial? - A Review of Mulsemedia Systems," *ACM Comput. Surv.*, vol. 51, no. 5, pp. 1–35, Sep. 2018, doi: 10.1145/3233774.

[20] Gartner, "Forecast unit shipments of wearable devices worldwide from 2017 to 2019 and in 2022 (in million units), by category," *Statista - The Statistics Portal*, 2018. https://www-statista-com.proxy.lib.sfu.ca/statistics/385658/electronic-wearable-fitness-devices-worldwide-shipments/ (accessed Mar. 14, 2019).

[21] Xania News, "Value of the social online games market worldwide from 2010 to 2020, by type (in million U.S. dollars)," *Statista - The Statistics Portal*. https://www.statista.com/statistics/558350/value-social-online-games-by-type-global/ (accessed Mar. 14, 2019).

[22] SuperData Research, "Immersive technology consumer market revenue worldwide from 2018 to 2022, by segment (in billion U.S. dollars)," *Statista - The Statistics Portal*. https://www.statista.com/statistics/936078/worldwide-consumer-immersive-technology-market-revenue/ (accessed Jun. 04, 2019).

[23] G. N. Yannakakis and J. Togelius, "A Panorama of Artificial and Computational Intelligence in Games," *IEEE Trans. Comput. Intell. AI Games*, vol. 7, no. 4, pp. 317–335, Dec. 2015, doi: 10.1109/TCIAIG.2014.2339221.

[24] A. Summerville *et al.*, "Procedural Content Generation via Machine Learning (PCGML)," *IEEE Trans. Games*, vol. 10, no. 3, pp. 257–270, 2017, doi: 10.1109/TG.2018.2846639.

[25] B. Kybartas and R. Bidarra, "A Survey on Story Generation Techniques for Authoring Computational Narratives," *IEEE Trans. Comput. Intell. AI Games*, vol. 9, no. 3, pp. 239–253, 2017, doi: 10.1109/TCIAIG.2016.2546063.

[26] M. V Barbera, S. Kosta, A. Mei, and J. Stefa, "To offload or not to offload? The bandwidth and energy costs of mobile cloud computing," in *2013 Proceedings IEEE INFOCOM*, Apr. 2013, pp. 1285–1293, doi: 10.1109/INFCOM.2013.6566921.

[27] J. Guo and Z. Gong, "Measuring virtual wealth in virtual worlds," *Inf. Technol. Manag.*, vol. 12, no. 2, pp. 121–135, Jun. 2011, doi: 10.1007/s10799-011-0082-9.

[28] S. Hetcher, "User-Generated Content and the Future of Copyright: Part Two - Agreements between Users and Mega-Sites Symposium Review," *St. Cl. Comput. High Technol. Law J.*, vol. 24, no. 4, pp. 829–868, 2007, [Online]. Available: https://heinonline.org/HOL/P?h=hein.journals/sccj24&i=839.

[29] D. Terry, "Replicated Data Consistency Explained Through Baseball," *Commun. ACM*, vol. 56, no. 12, pp. 82–89, Dec. 2013, doi: 10.1145/2500500.

[30] M. Mauve, J. Vogel, V. Hilt, and W. Effelsberg, "Local-lag and timewarp: providing consistency for replicated continuous applications," *IEEE Trans. Multimed.*, vol. 6, no. 1, pp. 47–57, Feb. 2004, doi: 10.1109/TMM.2003.819751.

[31] A. M. Khan, S. Chabridon, and A. Beugnard, "Synchronization Medium: A Consistency Maintenance Component for Mobile Multiplayer Games," in *Proceedings of the 6th ACM SIGCOMM Workshop on Network and System Support for Games*, 2007, pp. 99–104, doi: 10.1145/1326257.1326275.

[32] J. R. Millar, D. D. Hodson, G. L. Peterson, and D. K. Ahner, "Consistency and fairness in real-time distributed virtual environments: Paradigms and relationships," *J. Simul.*, vol. 11, no. 3, pp. 295–302, 2017, doi: 10.1057/s41273-016-0035-8.

[33] A. B. Bondi, "Characteristics of scalability and their impact on performance," in

*Proceedings of the 2nd international workshop on Software and performance*, 2000, pp. 195–203.

[34] H. Liu, M. Bowman, and F. Chang, "Survey of state melding in virtual worlds," *ACM Comput. Surv.*, vol. 44, no. 4, pp. 1–25, Sep. 2012, doi: 10.1145/2333112.2333116.

[35] C. E. B. Bezerra and C. F. R. Geyer, "A load balancing scheme for massively multiplayer online games," *Multimed. Tools Appl.*, vol. 45, no. 1, pp. 263–289, Oct. 2009, doi: 10.1007/s11042-009-0302-z.

[36] Irdeto, "Irdeto Global Gaming Survey: The Last Checkpoint for Cheating," 2018. [Online]. Available: https://www.statista.com/study/54578/cheating-in-video-gaming-report/.

[37] A. Yahyavi and B. Kemme, "Peer-to-peer architectures for massively multiplayer online games: A survey," *ACM Comput. Surv.*, vol. 46, no. 1, pp. 1–51, Jul. 2013, doi: 10.1145/2522968.2522977.

[38] E. Cuervo *et al.*, "MAUI: Making Smartphones Last Longer with Code Offload," in *Proceedings of the 8th International Conference on Mobile Systems, Applications, and Services*, 2010, pp. 49–62, doi: 10.1145/1814433.1814441.

[39] W. Cai *et al.*, "A survey on cloud gaming: Future of computer games," *IEEE Access*, vol. 4, pp. 7605–7620, 2016, doi: 10.1109/ACCESS.2016.2590500.

[40] A. P. Miettinen and J. K. Nurminen, "Energy Efficiency of Mobile Clients in Cloud Computing," in *2nd {USENIX} Workshop on Hot Topics in Cloud Computing, HotCloud'10, Boston, MA, USA, June 22, 2010*, 2010, [Online]. Available: https://www.usenix.org/conference/hotcloud-10/energy-efficiency-mobile-clients-cloud-computing.

[41] M. Altamimi, A. Abdrabou, K. Naik, and A. Nayak, "Energy Cost Models of Smartphones for Task Offloading to the Cloud," *IEEE Trans. Emerg. Top. Comput.*, vol. 3, no. 3, pp. 384–398, 2015, doi: 10.1109/TETC.2014.2387752.

[42] E. Brewer, "CAP twelve years later: How the 'rules' have changed," *Computer (Long. Beach. Calif).*, vol. 45, no. 2, pp. 23–29, Feb. 2012, doi: 10.1109/MC.2012.37.

[43] S. Gilbert and N. Lynch, "Perspectives on the CAP Theorem," *Computer (Long. Beach. Calif).*, vol. 45, no. 2, pp. 30–36, Feb. 2012, doi: 10.1109/MC.2011.389.

[44] D. Abadi, "Consistency Tradeoffs in Modern Distributed Database System Design: CAP is Only Part of the Story," *Computer (Long. Beach. Calif).*, vol. 45, no. 2, pp. 37–42, Feb. 2012, doi: 10.1109/MC.2012.33.

[45] K. J. L. Nevelsteen, "Survey of Pervasive Games and Technologies," in *A Survey of Characteristic Engine Features for Technology-Sustained Pervasive Games*, Cham: Springer International Publishing, 2015, pp. 11–39.

[46] I. Chatzigiannakis, G. Mylonas, O. Akribopoulos, M. Logaras, P. Kokkinos, and P. Spirakis, "The 'Hot Potato' Case: Challenges in Multiplayer Pervasive Games Based on Ad hoc Mobile Sensor Networks and the Experimental Evaluation of a Prototype Game," *arXiv e-prints*, p. arXiv:1002.1099, Feb. 2010, [Online]. Available: https://arxiv.org/abs/1002.1099.

[47] L. Fan, "Solving Key Design Issues for Massively Multiplayer Online Games on Peer-to-Peer Architectures," Heriot-Watt University, 2009.

[48] K. Zhang and B. Kemme, "Transaction Models for Massively Multiplayer Online Games," in *2011 IEEE 30th International Symposium on Reliable Distributed Systems*, Oct. 2011, pp. 31–40, doi: 10.1109/SRDS.2011.13.

[49] J. S. Gilmore and H. A. Engelbrecht, "A survey of state persistency in peer-to-peer massively multiplayer online games," *IEEE Trans. Parallel Distrib. Syst.*, vol. 23, no. 5, pp. 818–834, May 2012, doi: 10.1109/TPDS.2011.210.

[50] R. Nachiappan, B. Javadi, R. N. Calheiros, and K. M. Matawie, "Cloud storage reliability for Big Data applications: A state of the art survey," *J. Netw. Comput. Appl.*, vol. 97, pp. 35–47, 2017, doi: https://doi.org/10.1016/j.jnca.2017.08.011.

[51] A. Jøsang, R. Ismail, and C. Boyd, "A survey of trust and reputation systems for online service provision," *Decis. Support Syst.*, vol. 43, no. 2, pp. 618–644, 2007, doi: https://doi.org/10.1016/j.dss.2005.05.019.

[52] Z. Li, L. Liao, H. Leung, B. Li, and C. Li, "Evaluating the credibility of cloud services," *Comput. Electr. Eng.*, vol. 58, pp. 161–175, 2017, doi: 10.1016/j.compeleceng.2016.05.014.

[53] F. Bélanger and R. E. Crossler, "Privacy in the Digital Age: A Review of Information Privacy Research in Information Systems," *MIS Q.*, vol. 35, no. 4, pp. 1017–1042, Dec. 2011, [Online]. Available: http://dl.acm.org/citation.cfm?id=2208940.2208951.

[54] M. Theoharidou, N. Papanikolaou, S. Pearson, and D. Gritzalis, "Privacy Risk, Security, Accountability in the Cloud," in *2013 IEEE 5th International Conference on Cloud Computing Technology and Science*, Dec. 2013, vol. 1, pp. 177–184, doi: 10.1109/CloudCom.2013.31.

[55] S. Pearson, "Strong Accountability and Its Contribution to Trustworthy Data Handling in the Information Society," in *Trust Management XI*, 2017, pp. 199–218.

[56] F. Tietze, T. Pieper, and C. Herstatt, "To own or not to own: How ownership impacts user innovation–An empirical study," *Technovation*, vol. 38, pp. 50–63, 2015, doi: https://doi.org/10.1016/j.technovation.2014.11.001.

[57] P. Coutinho and R. José, "A Risk Management Framework for User-Generated Content on Public Display Systems," *Adv. Human-Computer Interact.*, vol. 2019, pp. 9769246:1-9769246:18, 2019, doi: 10.1155/2019/9769246.

[58] Y. Lin, B. Kemme, M. Patino-Martinez, and R. Jimenez-Peris, "Applying Database Replication to Multi-player Online Games," in *Proceedings of 5th ACM SIGCOMM Workshop on Network and System Support for Games*, 2006, doi: 10.1145/1230040.1230080.

[59] H. A. Engelbrecht and J. S. Gilmore, "Pithos: Distributed Storage for Massive Multi-User Virtual Environments," *ACM Trans. Multimed. Comput. Commun. Appl.*, vol. 13, no. 3, pp. 1–33, Jul. 2017, doi: 10.1145/3105577.

[60] E. Carlini, L. Ricci, and M. Coppola, "Flexible load distribution for hybrid distributed virtual environments," *Futur. Gener. Comput. Syst.*, vol. 29, no. 6, pp. 1561–1572, 2013, doi: 10.1016/j.future.2012.09.004.

[61]  B. Richerzhagen, D. Stingl, R. Hans, C. Gross, and R. Steinmetz, "Bypassing the cloud: Peer-assisted event dissemination for augmented reality games," in *14-th IEEE International Conference on Peer-to-Peer Computing*, 2014, pp. 1–10, doi: 10.1109/P2P.2014.6934296.

[62]  M. Varvello, C. Diot, and E. Biersack, "P2P second life: Experimental validation using Kad," in *Proceedings - IEEE INFOCOM*, Apr. 2009, pp. 1161–1169, doi: 10.1109/INFCOM.2009.5062029.

[63]  M. Amini and F. Osanloo, "Purpose-based Privacy Preserving Access Control for Secure Service Provision and Composition," *IEEE Trans. Serv. Comput.*, p. 1, 2018, doi: 10.1109/TSC.2016.2616875.

[64]  B. Shen, J. Guo, and Y. Yang, "MedChain: Efficient Healthcare Data Sharing via Blockchain," *Appl. Sci.*, vol. 9, no. 6, pp. 1–23, 2019, doi: 10.3390/app9061207.

[65]  M. Henze *et al.*, "Practical Data Compliance for Cloud Storage," in *2017 IEEE International Conference on Cloud Engineering (IC2E)*, Apr. 2017, pp. 252–258, doi: 10.1109/IC2E.2017.32.

[66]  M. Zhou, M. A. A. M. Leenders, and L. M. Cong, "Ownership in the virtual world and the implications for long-term user innovation success," *Technovation*, vol. 78, pp. 56–65, 2018, doi: 10.1016/j.technovation.2018.06.002.

[67]  J. Kim, H. Chung, M. Lee, Y. Chang, J. Jung, and P. Shin, "Design of user-centric semantic rights model for validation of user-generated content," *Cluster Comput.*, vol. 19, no. 3, pp. 1261–1273, Sep. 2016, doi: 10.1007/s10586-016-0578-5.

[68]  Z. Ma, M. Jiang, H. Gao, and Z. Wang, "Blockchain for digital rights management," *Futur. Gener. Comput. Syst.*, vol. 89, pp. 746–764, 2018, doi: https://doi.org/10.1016/j.future.2018.07.029.

[69]  W. Cai, M. Chen, and V. C. M. Leung, "Toward Gaming as a Service," *IEEE Internet Comput.*, vol. 18, no. 3, pp. 12–18, May 2014, doi: 10.1109/MIC.2014.22.

[70]  F. Chi, X. Wang, W. Cai, and V. C. M. Leung, "Ad-Hoc Cloudlet Based Cooperative Cloud Gaming," *IEEE Trans. Cloud Comput.*, vol. 6, no. 3, pp. 625–639, Jul. 2018, doi: 10.1109/TCC.2015.2498936.

[71]  M. Ghaffari, B. Hariri, S. Shirmohammadi, and D. T. Ahmed, "A Dynamic Networking Substrate for Distributed MMOGs," *IEEE Trans. Emerg. Top. Comput.*, vol. 3, no. 2, pp. 289–302, Jun. 2015, doi: 10.1109/TETC.2014.2330520.

[72]  R. Diaconu and J. Keller, "Kiwano: Scaling virtual worlds," in *Proceedings - Winter Simulation Conference*, Dec. 2017, pp. 1836–1847, doi: 10.1109/WSC.2016.7822230.

[73]  L. Ricci, L. Genovali, E. Carlini, and M. Coppola, "AOI-cast in distributed virtual environments: An approach based on delay tolerant reverse compass routing," *Concurr. Comput.* , vol. 27, no. 9, pp. 2329–2350, 2015, doi: 10.1002/cpe.2973.

[74]  C. Mönch, G. Grimen, and R. Midtstraum, "Protecting Online Games Against Cheating," in *Proceedings of 5th ACM SIGCOMM Workshop on Network and System Support for Games*, 2006, doi: 10.1145/1230040.1230087.

[75]  P. Laurens, R. F. Paige, P. J. Brooke, and H. Chivers, "A Novel Approach to the Detection

of Cheating in Multiplayer Online Games," in *12th IEEE International Conference on Engineering Complex Computer Systems (ICECCS 2007)*, Jul. 2007, pp. 97–106, doi: 10.1109/ICECCS.2007.11.

[76] J. Goodman and C. Verbrugge, "A Peer Auditing Scheme for Cheat Elimination in MMOGs," in *Proceedings of the 7th ACM SIGCOMM Workshop on Network and System Support for Games*, 2008, pp. 9–14, doi: 10.1145/1517494.1517496.

[77] S. Schuster and T. Weis, "Enforcing Game Rules in Untrusted P2P-based MMVEs," in *Proceedings of the 4th International ICST Conference on Simulation Tools and Techniques*, 2011, pp. 288–295, [Online]. Available: http://dl.acm.org/citation.cfm?id=2151054.2151106.

[78] S. Webb, S. Soh, and W. Lau, "RACS: a referee anti-cheat scheme for P2P gaming," in *Proceedings of the 17th international workshop on Network and operating systems support for digital audio and video*, 2007, pp. 34–42, [Online]. Available: http://hdl.handle.net/20.500.11937/32304.

[79] H. Flores, P. Hui, S. Tarkoma, Y. Li, S. Srirama, and R. Buyya, "Mobile code offloading: from concept to practice and beyond," *IEEE Commun. Mag.*, vol. 53, no. 3, pp. 80–88, Mar. 2015, doi: 10.1109/MCOM.2015.7060486.

[80] S. Shi and C.-H. Hsu, "A Survey of Interactive Remote Rendering Systems," *ACM Comput. Surv.*, vol. 47, no. 4, pp. 1–29, May 2015, doi: 10.1145/2719921.

[81] Y. Lin and H. Shen, "CloudFog: Leveraging Fog to Extend Cloud Gaming for Thin-Client MMOG with High Quality of Service," *IEEE Trans. Parallel Distrib. Syst.*, vol. 28, no. 2, pp. 431–445, Feb. 2017, doi: 10.1109/TPDS.2016.2563428.

[82] M. H. Gunes, M. Yuksel, and H. Ceker, "A blind processing framework to facilitate openness in smart grid communications," *Comput. Networks*, vol. 86, pp. 14–26, 2015, doi: https://doi.org/10.1016/j.comnet.2015.05.004.

[83] H. Wang, "Anonymous Data Sharing Scheme in Public Cloud and Its Application in E-Health Record," *IEEE Access*, vol. 6, pp. 27818–27826, 2018, doi: 10.1109/ACCESS.2018.2838095.

[84] J. Li, X. Lin, Y. Zhang, and J. Han, "KSF-OABE: Outsourced Attribute-Based Encryption with Keyword Search Function for Cloud Storage," *IEEE Trans. Serv. Comput.*, vol. 10, no. 5, pp. 715–725, 2017, doi: 10.1109/TSC.2016.2542813.

[85] B. Wang, B. Li, and H. Li, "Oruta: privacy-preserving public auditing for shared data in the cloud," *IEEE Trans. Cloud Comput.*, vol. 2, no. 1, pp. 43–56, Jan. 2014, doi: 10.1109/TCC.2014.2299807.

[86] S. J. Stolfo, M. B. Salem, and A. D. Keromytis, "Fog Computing: Mitigating Insider Data Theft Attacks in the Cloud," in *2012 IEEE Symposium on Security and Privacy Workshops*, May 2012, pp. 125–128, doi: 10.1109/SPW.2012.19.

[87] C. Modi, D. Patel, B. Borisaniya, A. Patel, and M. Rajarajan, "A survey on security issues and solutions at different layers of Cloud computing," *J. Supercomput.*, vol. 63, no. 2, pp. 561–592, Feb. 2013, doi: 10.1007/s11227-012-0831-5.

[88] R. Sarathy and C. J. Robertson, "Strategic and Ethical Considerations in Managing Digital Privacy," *J. Bus. Ethics*, vol. 46, no. 2, pp. 111–126, Aug. 2003, doi:

10.1023/A:1025001627419.

[89]  C. Dougherty, "Bragg v. Linden: Virtual Property Rights Litigation," *E-Commerce Law Policy*, vol. 9, no. 7, pp. 1–5, 2007, [Online]. Available: http://papers.ssrn.com/sol3/papers.cfm?abstract_id=1092284.

[90]  D. M. Ryu, *Evans v. Linden Research, Inc.*, vol. C-11–01078. 2014, pp. 1–9.

[91]  C. George and J. Scerri, "Web 2.0 and user-generated content: legal challenges in the new frontier," *J. information, law Technol.*, vol. 12, no. 2, pp. 1–22, 2007, [Online]. Available: http://www2.warwick.ac.uk/fac/soc/law/elj/jilt/2007_2/george_scerri/george_scerri.pdf.

[92]  B. Knutsson, Honghui Lu, Wei Xu, and B. Hopkins, "Peer-to-peer support for massively multiplayer games," in *IEEE INFOCOM 2004*, Mar. 2004, vol. 1, pp. 96–107, doi: 10.1109/infcom.2004.1354485.

[93]  E. Buyukkaya, M. Abdallah, and G. Simon, "A survey of peer-to-peer overlay approaches for networked virtual environments," *Peer-to-Peer Netw. Appl.*, vol. 8, no. 2, pp. 276–300, Mar. 2013, doi: 10.1007/s12083-013-0231-5.

[94]  R. Bhagwan, K. Tati, Y. C. Cheng, S. Savage, and G. M. Voelker, "Total recall: System support for automated availability management," in *Proc NSDI*, 2004, pp. 25–25, [Online]. Available: http://portal.acm.org/citation.cfm?id=1251175.1251200.

[95]  B. Shen and J. Guo, "Virtual Net: A Decentralized Architecture for Interaction in Mobile Virtual Worlds," *Wirel. Commun. Mob. Comput.*, vol. 2018, pp. 9749187:1-9749187:24, 2018, doi: 10.1155/2018/9749187.

[96]  S. Y. Hu, T. H. Huang, S. C. Chang, W. L. Sung, J. R. Jiang, and B. Y. Chen, "FLoD: A framework for peer-to-peer 3D streaming," in *Proceedings - IEEE INFOCOM*, Apr. 2008, pp. 2047–2055, doi: 10.1109/INFOCOM.2007.195.

[97]  Ş. B. Çevikbaş and V. İşler, "Phaneros: Visibility-based framework for massive peer-to-peer virtual environments," *Comput. Animat. Virtual Worlds*, vol. 30, no. 1, p. e1808, 2019, doi: 10.1002/cav.1808.

[98]  J. P. Cruz, Y. Kaji, and N. Yanai, "RBAC-SC: Role-Based Access Control Using Smart Contract," *IEEE Access*, vol. 6, pp. 12240–12251, 2018, doi: 10.1109/ACCESS.2018.2812844.

[99]  J. Lee, "BIDaaS: Blockchain Based ID As a Service," *IEEE Access*, vol. 6, pp. 2274–2278, 2018, doi: 10.1109/ACCESS.2017.2782733.

[100] A. Reyna, C. Martín, J. Chen, E. Soler, and M. Díaz, "On blockchain and its integration with IoT. Challenges and opportunities," *Futur. Gener. Comput. Syst.*, vol. 88, pp. 173–190, 2018, doi: https://doi.org/10.1016/j.future.2018.05.046.

[101] R. Neisse, G. Steri, and I. Nai-Fovino, "A Blockchain-based Approach for Data Accountability and Provenance Tracking," in *Proceedings of the 12th International Conference on Availability, Reliability and Security*, 2017, pp. 14:1--14:10, doi: 10.1145/3098954.3098958.

[102] M. Vogel, "Final Fantasy XV Pocket Edition HD Review," *Nintendo Life*, 2018. http://www.nintendolife.com/reviews/switch-eshop/final_fantasy_xv_pocket_edition_hd (accessed May 26, 2019).

[103] N. Gillham, "Final Fantasy XV: Pocket Edition review," *GodisaGeek*, 2018. https://www.godisageek.com/reviews/final-fantasy-xv-pocket-edition-review/ (accessed May 26, 2019).

[104] E. Carlini, M. Coppola, and L. Ricci, "Integration of P2P and clouds to support massively multiuser virtual environments," in *2010 9th Annual Workshop on Network and Systems Support for Games, NetGames 2010*, Nov. 2010, pp. 1–6, doi: 10.1109/NETGAMES.2010.5679660.

[105] M. Flintham *et al.*, "Where On-line Meets on the Streets: Experiences with Mobile Mixed Reality Games," in *Proceedings of the SIGCHI Conference on Human Factors in Computing Systems*, 2003, pp. 569–576, doi: 10.1145/642611.642710.

[106] A. Bujari, L. De Giovanni, and C. E. Palazzi, "Optimal configuration of active and backup servers for augmented reality cooperative games," *Concurr. Comput. Pract. Exp.*, vol. 30, no. 20, p. e4454, 2018, doi: 10.1002/cpe.4454.

[107] W. Cai, V. C. M. Leung, and L. Hu, "A Cloudlet-Assisted Multiplayer Cloud Gaming System," *Mob. Networks Appl.*, vol. 19, no. 2, pp. 144–152, Apr. 2014, doi: 10.1007/s11036-013-0485-4.

[108] M. T. Najaran and C. Krasic, "Scaling online games with adaptive interest management in the cloud," in *2010 9th Annual Workshop on Network and Systems Support for Games*, Nov. 2010, pp. 1–6, doi: 10.1109/NETGAMES.2010.5680282.

[109] C. Carter, A. E. Rhalibi, M. Merabti, and A. T. Bendiab, "Hybrid Client-Server, Peer-to-Peer framework for MMOG," in *2010 IEEE International Conference on Multimedia and Expo*, Jul. 2010, pp. 1558–1563, doi: 10.1109/ICME.2010.5583228.

[110] D. T. Ahmed, S. Shirmohammadi, and J. C. De Oliveira, "A hybrid P2P communications architecture for zonal MMOGs," *Multimed. Tools Appl.*, vol. 45, no. 1–3, pp. 313–345, Oct. 2009, doi: 10.1007/s11042-009-0311-y.

[111] E. Carlini, L. Ricci, and M. Coppola, "Integrating centralized and peer-to-peer architectures to support interest management in massively multiplayer on-line games," *Concurr. Comput.*, vol. 27, no. 13, pp. 3362–3382, 2015, doi: 10.1002/cpe.3289.

[112] H. Kavalionak, E. Carlini, L. Ricci, A. Montresor, and M. Coppola, "Integrating peer-to-peer and cloud computing for massively multiuser online games," *Peer-to-Peer Netw. Appl.*, vol. 8, no. 2, pp. 301–319, Mar. 2015, doi: 10.1007/s12083-013-0232-4.

[113] E. K. Lua, J. Crowcroft, M. Pias, R. Sharma, and S. Lim, "A survey and comparison of peer-to-peer overlay network schemes," *IEEE Commun. Surv. Tutorials*, vol. 7, no. 2, pp. 72–93, 2005, doi: 10.1109/COMST.2005.1610546.

[114] Flexera Software, "Challenges of using cloud computing worldwide as of 2019," *Statista - The Statistics Portal*. https://www.statista.com/statistics/511283/worldwide-survey-cloud-computing-risks/ (accessed Jun. 16, 2019).

[115] N. Oualha, J. Leneutre, and Y. Roudier, "Verifying remote data integrity in peer-to-peer data storage: A comprehensive survey of protocols," *Peer-to-Peer Netw. Appl.*, vol. 5, no. 3, pp. 231–243, Sep. 2012, doi: 10.1007/s12083-011-0117-3.

[116] K. Matsumoto and Y. Okabe, "A Collusion-Resilient Hybrid P2P Framework for Massively Multiplayer Online Games," in *2017 IEEE 41st Annual Computer Software and*

*Applications Conference (COMPSAC)*, Jul. 2017, vol. 2, pp. 342–347, doi: 10.1109/COMPSAC.2017.10.

[117] B. Shen and J. Guo, "Efficient Peer-to-peer Content Sharing for Learning in Virtual Worlds," *J. Univers. Comput. Sci.*, vol. 25, no. 5, pp. 465–488, 2019.

[118] B. Shen, J. Guo, and L. X. Li, "Cost optimization in persistent virtual world design," *Inf. Technol. Manag.*, vol. 19, no. 3, pp. 155–169, Sep. 2018, doi: 10.1007/s10799-017-0283-y.
30